# Vibrational Stark Spectroscopy on Fluorobenzene with Quantum Cascade Laser Dual Frequency Combs

Urszula Szczepaniak[1,i], Samuel H. Schneider[2], Raphael Horvath[1], Jacek Kozuch[2], Markus Geiser[1]

1 IRsweep AG, Laubisruetistrasse 44, 8712 Staefa, Switzerland

2 Department of Chemistry, Stanford University, Stanford, California 94305-5012

## Abstract

We demonstrate the performance of a dual frequency comb QCL spectrometer for the application of vibrational Stark spectroscopy. Measurements performed on fluorobenzene with the dual-comb spectrometer (DCS) were compared to results obtained using a conventional Fourier transform infrared (FTIR) instrument in terms of spectral response, parameter estimation, and signal-to-noise ratio. The dual-comb spectrometer provided similar qualitative and quantitative data as the FTIR setup in 250 times shorter acquisition time. For fluorobenzene, the DCS measurement resulted in a more precise estimation of the fluorobenzene Stark tuning rate ((0.81 ± 0.09) cm-1/(MV/cm)) than with the FTIR system ((0.89 ± 0.15) cm-1/(MV/cm)). Both values are in accordance with the previously reported value of 0.84 cm-1/(MV/cm). We also point to an improvement of signal to noise ratio (SNR) in the DCS configuration. Additional characteristics of the dual-comb spectrometer applicable to vibrational Stark spectroscopy and their scaling properties for future applications are discussed.

## Introduction

The influence of electric fields on optical spectra is known as the Stark effect or electrochromism.[1] While it has been extensively exploited in electronic Stark spectroscopy[2–6], similar effects can also be observed in vibrational spectra[7,8], i.e. the vibrational Stark effect (VSE), where an electric field perturbs a vibrational mode's ground and excited states, resulting in a shift of its absorption energy. This effect has been used in the study of proteins[9–11] and other biological systems[12–14], electrode interfaces[15–17], solute-solvent interactions[18] etc., providing insight into the nature of electrostatics on the molecular level, a topic of general importance in biology, chemistry, and materials science[19]. In this way, it has contributed towards understanding the effect of electric fields on the anharmonicity of

---

i Corresponding Author: urszula.szczepaniak@irsweep.com

chemical bonds, the band structure in materials, binding and catalytic processes, as well as on the transition state stabilization in enzymes; the latter is particularly relevant in the field of protein design and engineering, and its applications in biocatalysis.[19]

The framework of the VSE enables quantification of the magnitude of electric fields or their changes using suitably calibrated vibrational probes, e.g. local high frequency modes such as carbonyls and nitriles. Vibrational Stark spectroscopy (VSS) is an experimental approach to directly measure the sensitivity of a vibrational mode to an external electric field, providing such a quantitative calibration for inferring electric fields in condensed phase systems using vibrational spectroscopy (e.g. Fourier-Transform Infrared (FTIR), Raman, 2D-IR, etc.).[7,20–24] It is generally performed in an isotropic frozen glass, where the change in absorbance in the presence and absence of an electric field can be related to the derivatives of the absorbance spectrum in the absence of an external field. High frequency local modes are generally observed to follow the linear Stark effect[8]. That is, their response to an electric field arises predominantly from the difference dipole, referred to as the Stark tuning rate ($\Delta\mu = \mu_1 - \mu_0$; units of $cm^{-1}/(MV/cm)$), between the ground and first-excited states of the vibrational mode. In VSS, analysis of the 2nd-derivative contribution to the observed field-on minus field-off spectrum can be directly used to quantify the Stark tuning rate. Usually, the VSS spectra in the mid-IR region are probed with the use of Fourier transform infrared (FTIR) instruments.[19] While suitable for a variety of possible applications, the approach is currently limited to the analysis of small molecules with high solubility (ca. ≥100 mM) in an appropriate frozen polymer or glassy matrix where high signal-to-noise ratios (SNR) can be achieved. This limitation is brought about mainly by the low brightness of commonly used IR light sources in FTIR spectrometers in combination with the low sensitivity of the VSS signal, the low oscillator strength (in comparison to many electronic transitions) and the isotropy of the frozen sample. Consequently, the low brightness of FTIR globars naturally results in long experimental durations, which can potentially increase the likelihood of dielectric breakdown from the externally applied electric field. Therefore, a rapid technique with a high-power polychromatic source in the mid-IR is sought.

One way to overcome the low brightness of globar sources could be the use of an FTIR equipped with a coherent light source, such as a femtosecond laser.[25,26] Such a solution offers very fast spectral collection (on the order of 10 s with very high SNR), but its use is limited to the near-IR spectral range. Replacement of the FTIR globar with a QCL has been demonstrated and would combine the benefits of the brighter light source with the established spectrometers.[27,28] On the other hand, also EC-QCL systems are benchmarked with FTIR.[29] A limitation of laser-based systems would be a spectral bandwidth much narrower than that covered by FTIR, which can necessitate tuning or exchanging of the lasers if different spectral ranges are to be monitored.

Recent advances in Dual-Comb Spectrometers (DCS) [30–32] including mid-IR DCS[33–36] present the technique as a powerful molecule identification tool in mid-IR spectroscopy that finds use in demanding applications[37]. Generating frequency combs in the mid-IR region is, however, a challenging task.[38] Thanks to the discovery of a semiconductor QCL-based electrically pumped frequency-comb source[39], dual-comb mid-IR spectroscopy systems were substantially simplified.[40–42]

We present VSS measurements on a fluorobenzene sample using a new high-brightness, high-speed, and multi-wavelength DCS based on quantum cascade laser (QCL) frequency combs in the $1170 - 1230\ cm^{-1}$ range. The spectral spacing of comb modes of $0.328\ cm^{-1}$ exceeds the requirements for studies of solids or fluids. The goal of this study was to benchmark a DCS instrument against the existing method currently in use for these spectroscopic studies and highlight differences between performance of the methods. Therefore, the results are compared to those obtained with a conventional FTIR spectrometer under identical sample setups (i.e. cryostat, sample-holder, optical windows and coatings, etc., as outlined in the experimental section). The data acquisition performed for FTIR was previously optimized for the VSS samples and is routinely used[7,43–45,10]. We also give an outlook regarding the expanded experimental scope accessible via DCS-based spectrometers for VSS.

# Experimental

## Sample Preparation

Solutions of 100 mM fluorobenzene (Sigma Aldrich, Inc.; see Figure 1-inset) in 2-methyltetrahydrofuran (ACROS Organics) were prepared for both FTIR and dual comb spectroscopy measurements. The inner surfaces of $CaF_2$ (1 mm thickness, 12.7 mm diameter, FOCtek Photonics) windows were coated with 4.5 nm of Ni, which acts as an electrode in the experiments. The sample was placed between two electrode-coated windows offset with ca. 25 µm Teflon spacers and immersed in home-built liquid nitrogen cryostat[46] equipped with IR-transparent windows to form an isotropic frozen glass. The capacitor thickness was determined using interferometry at room-temperature via UV-visible spectroscopy. The voltage was applied with a Trek 10/10 high-voltage power amplifier. The applied field was calculated as the ratio of the voltage applied to the sample and sample thickness, and thus it reflects the average macroscopic electric field across the sample capacitor. Further spectrometer-specific experimental details are outlined below.

## Dual Frequency-Comb Spectroscopy

Experiments were performed with a table-top dual frequency comb spectrometer (IRsweep IRis-F1).[47,48] The dual-comb system (Figure 1 – left) is based on two free-running quantum cascade laser (QCL) frequency combs that individually span over 70 cm$^{-1}$ and are centered at ~1220 cm$^{-1}$.[49] The combs overlap in the 1173 to 1230 cm$^{-1}$ region which results in dual-comb absorption spectra covering 57 cm$^{-1}$ with 0.328 cm$^{-1}$ point spacing. The average power per comb tooth is >2 mW with a total continuous wave output power of about 700 mW. The laser beams are combined, split into two beams on a 50:50 $CaF_2$ beam splitter and attenuated with neutral density filters. One of the combined beams is focused on a high-bandwidth HgCdTe reference detector, while the other passes through the sample (beam size of ca. 3 mm) and is then detected by a second identical HgCdTe detector.

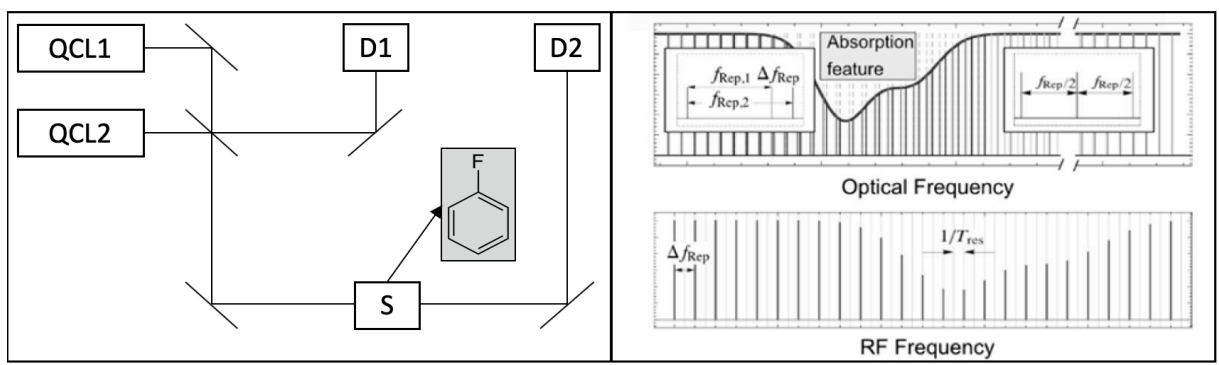

Figure 1. Left: Setup for dual-comb spectroscopy. Two quantum cascade lasers (QCL1, QCL2) generate frequency combs that travel (i) to a fast AC-coupled HgCdTe detector (D1) (ii) through sample compartment (S) with an optically active sample (flurobenzene - inset) to an identical HgCdTe detector (D2). Right: Top panel: Representation of two frequency combs with slightly differing repetition frequencies ($f_{Rep,1}$, $f_{Rep,2}$). Bottom: Beating signal of the interleaving combs shown in the top panel. Information from optical range is mapped onto the radio frequency range. Adapted from Ref. [47].

Overlapping two frequency combs allows simultaneous detection of the optical modes of the lasers due to a multi-heterodyne detection scheme. The mixing of two frequency-combs (Figure 1 – right top) results in a heterodyne beat spectrum with the frequency spacing between beatings of neighboring comb teeth in the radio frequency range (Figure 1 – right bottom): $\Delta f_{Rep} = f_{Rep,1} - f_{Rep,2}$, where $f_{Rep,1}$ and $f_{Rep,2}$ are repetition frequencies of the lasers 1 and 2, respectively. The achievable time resolution, connected with resolving neighboring comb teeth, is limited by $\Delta f_{Rep}$, according to the relation: $T_{res} = 1/\Delta f_{Rep}$. Due to high repetition rate of the used QCLs, enabled by a low cavity length, short acquisition time and high time resolution (<1 μs) can be achieved.[47] Total power on the detector was observed with a DC-coupled port of the detector, ensuring operation in a linear regime.

Figure 2A shows the different triggering schemes of the experiment, resulting in different cycle times for the two methods. In the dual-comb experiment, the voltage was controlled with a trigger; data was acquired for 16 ms, consisting of 8 ms with voltage on and 8 ms with voltage off. This *acquisition time per cycle*, where one cycle refers to one acquired difference spectrum, multiplied by the number of acquisitions, is later referred to as *the acquisition time*. The trigger signals were separated in time by 1 s, yielding 1 s cycles (i.e. *experimental time per cycle*) and defining *the experiment time* as multiplication of number of acquisitions by 1 s. This acquisition scheme resulted in 984 ms experimental dead time. About half of that time was used for data saving, while the other half was waiting time for the trigger. This scheme was chosen at the time of the experiments since the DCS was optimized for single shot[ii] experiments reported previously.[47,48]

The Stark spectra are obtained by calculating the difference in sample absorption in the presence and absence of an externally applied electric field with up to 96 spectra being averaged. Two sets of voltages were applied: 2.0 kV, and 3.0 kV, to a 28.0 ± 0.1 μm thick sample, corresponding to fields of 0.715 MV/cm, and 1.073 MV/cm, respectively. It was verified that the Stark signal scales with the square of the electric field intensity.[19]

---

[ii] In this case, "single shot" refers to a single acquisition consisting of a continuous train of spectra that correspond to a unique event, as opposed to "multiple shot", which would require experimental repetitions.

The low-temperature absorbance spectrum required for Stark analysis was measured via FTIR and further used for wavenumber calibration of the DCS spectrum.

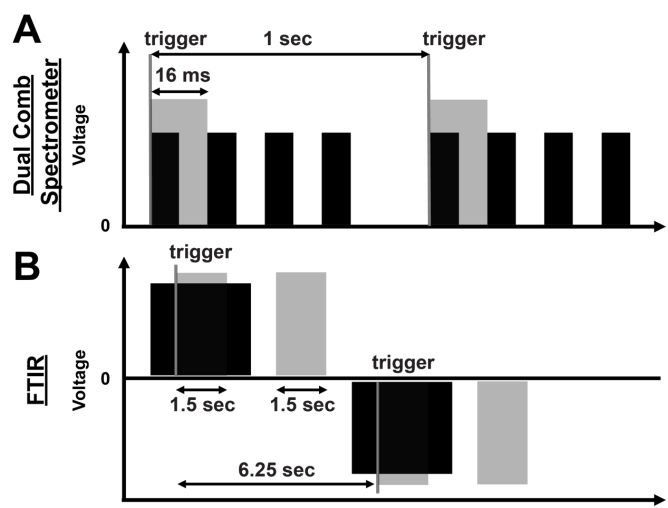

Figure 2. Schematic of a triggered experiment. (A) Dual-comb spectrometer. (B) FTIR. Black areas indicate time when the voltage was applied, grey areas and corresponding numbers indicate acquisition time. Time between the two vertical lines, denoted by "trigger", indicates experimental time per cycle. Not drawn to scale, see text for further details.

# Fourier Transform Infrared Spectroscopy

The FTIR measurements were performed with a Bruker Vertex 70 FTIR spectrometer equipped with a globar source and a liquid nitrogen-cooled HgCdTe detector. Spectra were measured with 1.0 cm$^{-1}$ resolution and 128 scans (field on - field off) were averaged. A similar principle of taking difference spectra as for dual-comb spectrometer measurements was used here. The cycles were realized as presented in Figure 2B. The *experimental time per cycle* was 6.25 s and yielded a total *experiment time* of 800 s. Each cycle constituted an acquisition time of 3 s resulting from 1.5 s of data acquisition per field-on or field-off scan. The remaining 3.25 s was used to process and store the data. The applied voltage was 2.07 kV, and the sample thickness 27.78 ± 0.03 μm, yielding a field of 0.745 MV/cm. The experimental parameters for both systems are gathered in Table 1.

The resulting interferograms were transformed into absorbance spectra using a Blackman-Harris 3-Term apodization function in the range of 4000-1000 cm$^{-1}$ with a phase resolution of 32 and a zero-filling factor of 2. In general, the VSS measurements performed with conventional FTIR instruments are photon-limited using the globar source due to the Ni-coating and strongly-absorbing frozen glasses. Therefore, the maximum J-stop of 6 mm is routinely used, and results in signals of ca. 1/3 of the maximum peak light intensity that can be detected using a liquid nitrogen-cooled HgCdTe detector.

To assure a well-defined absorption lineshape (both good SNR and accuracy in the wings required for further use of the derivatives; see later in text), the bulk absorption for both spectrometer analyses was measured in the same FTIR configuration after Stark spectra were recorded and is a potential source of deviation in the Stark tuning rate determination presented below. Immersing the sample in liquid N$_2$ results in an error in absorbance readout of about ±0.01 absorbance units.[19] The FTIR spectra were recorded and extracted with the use of commercially available software (Bruker OPUS 5.5).

Table 1. The auxiliary data for spectra obtained with DCS and FTIR instruments used for Data and Performance Analysis. One cycle is defined as 1 scan field-on followed by 1 scan field-off together with associated processing time before the next measurement with field on begins. Note that in the following discussion, only the acquisition times will be considered since experimental times are sample and application-dependent.

|  | Dual-comb spectrometer | FTIR |
|---|---|---|
| Exp. time per cycle | 1 s | 6.25 s |
| Acquisition time per cycle | 16 ms | 3 s |
| Number of cycles | 96 | 128 |
| Experiment time | 96 s | 800 s |
| Acquisition time | 1.536 s | 384 s |
| Applied field | 0.715 MV/cm | 0.745 MV/cm |
| Measured span | 1173-1230 cm$^{-1}$ | 1000-4000 cm$^{-1}$ |
| Resolution | 0.0003 cm$^{-1}$ | 1 cm$^{-1}$ |
| Point spacing | 0.328 cm$^{-1}$ | 0.5 cm$^{-1}$ |
| Spectral elements | 174 | 3000 |

## Data Analysis

In conducting the data analysis, we compare the Stark tuning rate from analytical weighted fitting between DCS and FTIR. For the comparison we used the Stark spectra measured at comparable electric fields, i.e. 0.715 and 0.745 MV/cm-1 in the case of the DCS and FTIR spectra, respectively, and reported spectra are subsequently shown scaled to a field strength of 1 MV/cm.

Optical power is unevenly distributed among the spectral elements of the DCS system, resulting in different variances of the individual spectral elements. This variance can be extracted for each spectral element individually. The inverse values are used as fitting weights, $w_i$, for DCS[iii] (see below and the details in the function description, available also in the SI).[47] Therefore, using "raw" DCS data with wavenumber-dependent standard deviation

---

[iii] i.e. more stable lines, showing lower noise, therefore higher SNR, are given more weight for a more reliable fit.

gives us the benefit of having additional information for the fit. For FTIR data, equal weights were assigned to all spectral elements.

The Stark tuning rate was independently determined for the DCS and FTIR data through a non-linear curve fitting procedure summarized below. The details of the fitting procedure and the corresponding parts of Python scripts used for the evaluation are shown in the SI. At first, the absorbance peak of fluorobenzene (A; recorded by FTIR and expressed in units of extinction coefficient, i.e. $M^{-1}$ $cm^{-1}$) was fit with a pseudo-Voigt profile yielding a function of wavenumber ($\tilde{v}$), i.e. $A(\tilde{v})$. The first and second derivatives of the quotient absorbance and the wavenumber $A(\tilde{v})/\tilde{v}$ were calculated. The Stark data (both DCS and FTIR in units of extinction coefficient) were normalized by the applied field and then independently fit with a weighted global fit to the curve: $\Delta A = aA(\tilde{v}) + b\frac{d}{d\tilde{v}}\frac{A(\tilde{v})}{\tilde{v}} + c\frac{d^2}{d^2\tilde{v}}\frac{A(\tilde{v})}{\tilde{v}}$. Coefficients $a$, $b$, and $c$ were found through a weighted non-linear least squares minimization (*curve_fit* function of Python library *scipy*). The fitting procedure yielded the value of $c$, as well as the covariance matrix from which the standard deviation of $c$ ($\sigma_c$) could be determined. The Stark tuning rate was obtained as, $\Delta\mu = \sqrt{10c}$[44] and its uncertainty (see Figure 4) was taken as $\sqrt{10\sigma_c}$.

It should be noted, that the fitting was performed on 0.0003 $cm^{-1}$ resolution data with 0.328 $cm^{-1}$ point spacing in the case of DCS and on 1 $cm^{-1}$ resolution with 0.5 $cm^{-1}$ point spacing data in the case of FTIR. I.e. for DCS, the very narrow laser bands are probing a broadband feature. More points in DCS spectrum could benefit the quality of the fit.

Another quantity that reflects the performance of both systems is the SNR. In the case of FTIR, the SNR evaluation procedure is well established. According to this conventional method, signal is divided by the noise, where the noise is determined in a part of spectrum without spectral features, assuming that the noise is equal throughout the spectrum. Such a procedure is not possible for DCS data since we are limited by the DCS coverage and recognize the occurrence of additional physical processes apart from the Stark response of the analyte that produce spectral features in the whole covered range.

Therefore, we use a similar approach, where the noise was assessed on the Stark peak of interest. To offer the most straightforward estimate of the SNR values, the SNR evaluation procedure was applied on the spectra that were processed to match the resolution between DCS and FTIR, as presented in Figure 3. The spectral range containing the Stark peak (1200-1230 $cm^{-1}$) was fitted (without baselining or weighting) for DCS and FTIR. The Stark peak was fit in both cases to the same analytical function. The rms of the residuals was then quantified for each system providing the *noise* value for the SNR evaluation, which is comparable to the spectrometer noise in the FTIR for back-to-back spectra (see Figure 6 bottom panel). The value of *signal* for SNR was obtained from fitted values as the difference between maximum and minimum value of the fit. This simple approach is only used to compare DCS and FTIR on equal footing. Instead, weighted fitting is recommended for other studies using DCS for the reasons outlined above.

# Results and Discussion

## Spectral Response

Both systems produced very similar spectral response before any processing (i.e. convolution, baselining, or fitting). In order to enable fair comparison between the DCS and FTIR spectrum, the former was convoluted using Blackmann-Harris filter to match the FTIR resolution (1 cm$^{-1}$, Figure 3). It can be seen, that both systems show almost identical Stark peaks, as well as a feature at 1180 cm$^{-1}$ that can be partially attributed to a response of the solvent. The greatest deviation between the two datasets is apparent around 1200 cm$^{-1}$, and this spectral region also corresponds to the greatest standard deviation in DCS signal, as is discussed below. The unconvoluted DCS spectra were then used to determine the Stark tuning rates (as discussed in the Data Analysis section).

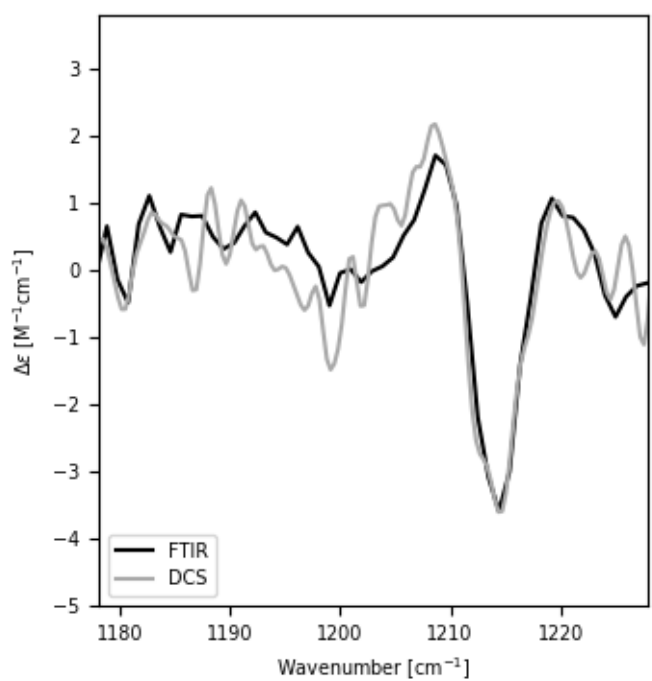

Figure 3. Comparison of the spectral response of FTIR[iv] and DCS instruments at 1 cm-1 resolution without baselining. The DCS spectrum was convoluted to match the resolution of FTIR. The spectra are shown scaled to a field strength of 1 MV/cm (experimental fields were 0.715 (DCS) and 0.745 (FTIR) MV/cm). Note that these spectra were used for the SNR evaluation according to the conventional method outlined in the Data Analysis section.

## Determination of the Stark Tuning Rate of Fluorobenzene

Data from the DCS and FTIR produce the same pronounced second-derivative-like Stark lineshape, as shown in Figure 3 and Figure 4, and are in accordance with previously reported spectra of fluorobenzene, centered at 1214.2 cm$^{-1}$.[45] The determined Stark tuning rate values ($|\Delta\mu|f$) are (0.81 ± 0.09) and (0.89 ± 0.15) cm$^{-1}$/(MV/cm) from the DCS and FTIR, respectively. Slight differences in the baselines or deviations in the sample thickness between room temperature and 77 K, at which the thickness was determined and the VSS experiment was performed, may contribute to the deviation in Stark tuning rate between FTIR and DCS. According to Bublitz and Boxer[19] the latter can result in errors of ± 1-2 μm. Nevertheless, both values are in accordance within the determined error margins of that

---

[iv] Full FTIR spectrum is available in the SI.

previously measured[45] for the C-F stretch of fluorobenzene in 2-methyltetrahydrofuran, i.e. $|\Delta\mu|f$ = 0.84 cm$^{-1}$/(MV/cm).

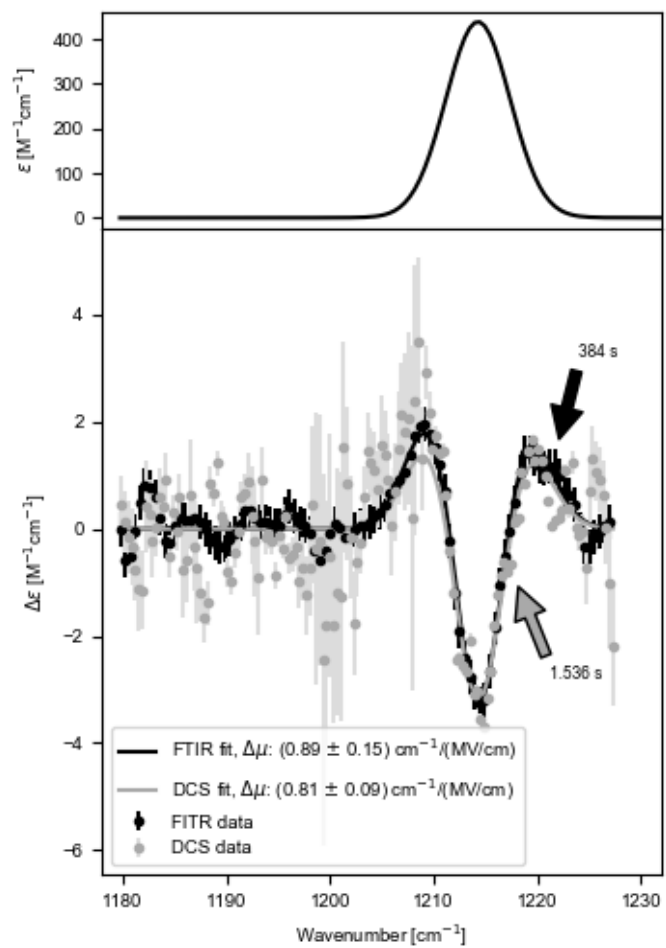

Figure 4. Top: Pseudo-Voigt fit of the IR absorption spectrum of 100 mM fluorobenzene in glassy 2-MeTHF at 77K. Bottom: Stark spectra scaled to an electric field strength of 1 MV/cm as registered with dual-comb spectrometer (grey, after total acquisition time of 1.536 s, equivalent to 96 cycles with a spectral point spacing of 0.328 cm$^{-1}$) and FTIR (black, after total acquisition time of 384 s, equivalent to 128 cycles at a spectral resolution of 1 cm$^{-1}$). Error bars correspond to standard deviation of the laser line intensity (DCS) and RMSE of the linear fit determined for the part of the spectrum below 1200 cm$^{-1}$ (FTIR). Note parts of the DCS spectra with low noise levels. Fits using a linear combination of absorbance derivatives are shown, which were used to determine $\Delta\mu$ and the SNR (see text for details).

The DCS spectra presented in Figure 4 were recorded with an acquisition time of 1.536 s. Even shorter times can be sufficient to obtain the information of interest. In order to test the sensitivity of the set-up we compared the spectra for various acquisition times. Figure 5 presents Stark spectra as registered for 8 to 96 averages, corresponding to 128 ms to 1.536 s acquisition time, respectively. The Stark response can be observed with acquisition times as short as 128 ms.

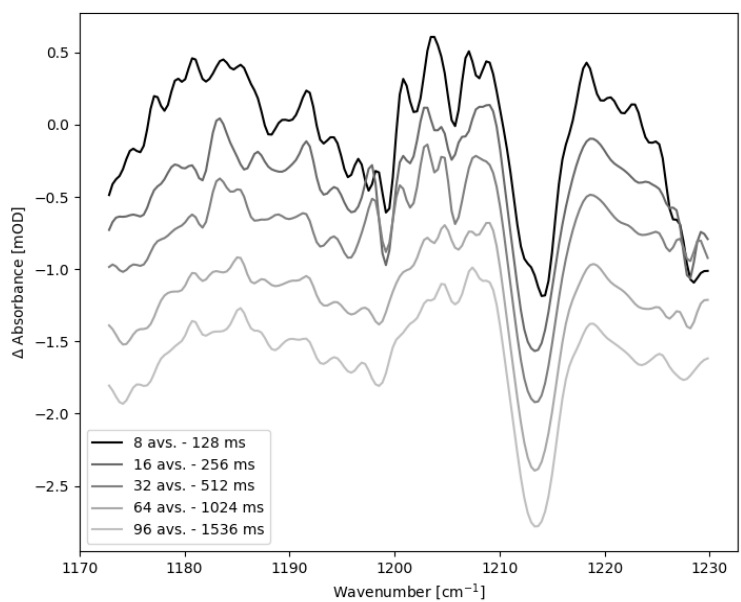

Figure 5. Stark spectra obtained with DCS with increasing averaging and corresponding acquisition times, measured at an external electric field of 1.07 MV/cm and with a spectral point spacing of 0.328 cm⁻¹. No baselining was applied. An offset of 0.4 mOD per spectrum was used for better visualization of the curves.

## Performance Analysis

In the previous sections, we have shown that both techniques provide similar spectral responses and Stark tuning rates. Quantifying the SNR for both spectra in Figure 3 (according to the method as outlined in the Data Analysis section and Figure 6) results in an estimation of SNR of 9.5 and 20.2 for the DCS and FTIR, respectively. In order to fairly compare the SNR values between DCS and FTIR, we further normalize these ratios based on the acquisition times (i.e. 1.536 s or 384 s, respectively) and applied fields (of 0.715 and 0.745 MV/cm; note that the signal scales with field squared). The resulting corrected ratio of SNR values is:

$$\left(\frac{SNR_{DCS}}{SNR_{FTIR}}\right)_{corr} = \frac{SNR_{DCS}}{SNR_{FTIR}} \cdot \frac{\sqrt{\text{time } FTIR}}{\sqrt{\text{time } DCS}} \cdot \left(\frac{F_{FTIR}}{F_{DCS}}\right)^2 = \frac{SNR_{DCS}}{SNR_{FTIR}} \cdot \frac{\sqrt{384 \text{ s}}}{\sqrt{1.536 \text{ s}}} \cdot \left(\frac{0.745\frac{MV}{cm}}{0.715\frac{MV}{cm}}\right)^2 \approx 8 \text{ (Eq. 1)}$$

Accounting for the acquisition times and electric fields, we determine that the SNR for the DCS is improved by a factor of ca. 8 relative to the FTIR.

Note that this evaluation assumes that the duty cycles are equivalent, which may vary depending on spectrometer and specific application, such as implementation of triggering, data acquisition, and data recording. At the time of the experiments the duty cycles were 1.6% (we applied setting from previous work in reference 47) and 48% for the DCS and FTIR spectrometers, respectively. Currently, for the DCS system, the duty cycle reaches ca. 20-26%, depending on the exact acquisition scheme and performance of the computer hardware.

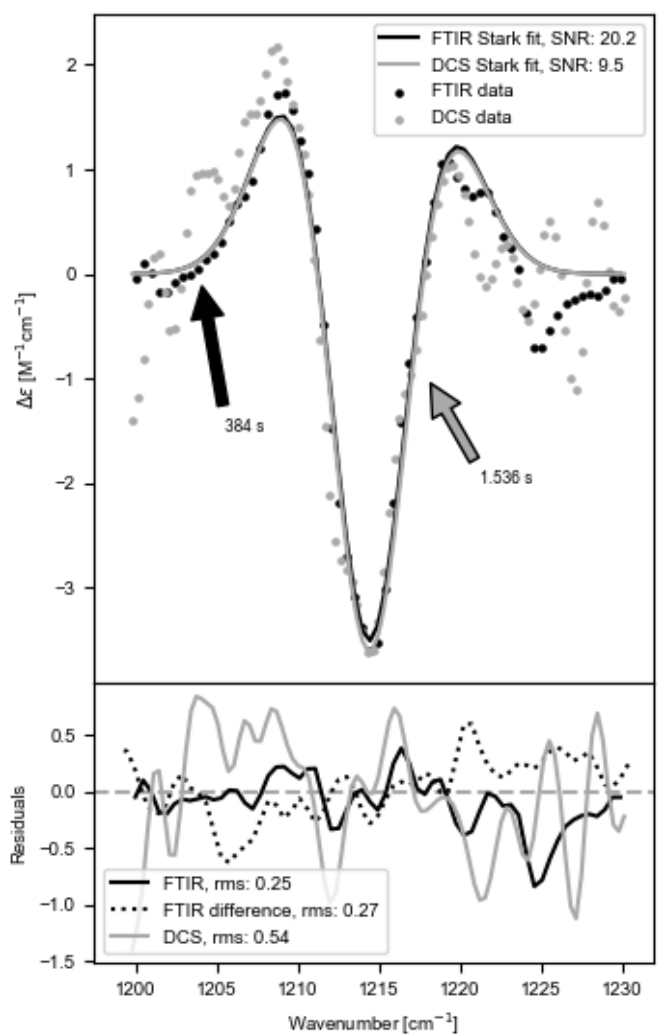

Figure 6. Residual evaluation for FTIR and DCS spectra. Top: Stark peak from the spectra as presented in Figure 3 with their unweighted fits to the sum of fluorobenzene absorbance derivatives (see text for details). No baselining was applied. Bottom: The residuals of the fits (solid lines). For comparison, the difference between two consecutive FTIR spectra measured at low temperature only with 2MeTHF in the cryostat (i.e. without the sample responsible for the spectral feature in this region, but otherwise same conditions as in the Stark spectrum) is presented (dotted line).

In order to account for the wavelength-dependent noise of the DCS system, we outline other approaches that can be taken to evaluate the SNR. Note that there are well established approaches for DCS setups, such as one outlined in Ref. 50 developed for an instrument-dependent (and not sample-dependent) comparison. It accounts for the SNR, acquisition time, and number of resolved frequency elements. There are however a number of assumptions that do not hold in the present case, such as, homogenous intensity distribution, weak absorptions, and an asymmetric configuration in which only one comb passes through the sample. While it is worthwhile to include these parameters in a theoretical framework for noise comparison of comb systems, this is beyond the scope of this paper. Moreover, for the application of VSS, i.e. determination of the Stark tuning rate from analytical fitting of a single sharp peak, the SNR value used above gives an application-relevant metric for the comparison. Despite the improved SNR using the DCS system, we acknowledge that the FTIR offers a much broader spectral range (Table 1) relative to the DCS, requiring multiple lasers to achieve comparable coverage.

In order to quantify the noise specific to the DCS, we measure the standard deviation of the transmission signal at each spectral element with a reference detector. This is demonstrated as the error bars in Figure 4 (and can be also visualized in Figure 7S in the form of a histogram). Although it is not strictly corresponding to the usual definition of spectral noise, it reflects the measurement stability at each spectral element. Hence, one could also consider taking a whole "spectrum" of the standard deviation of the transmitted signal as the noise for evaluating the performance of the DCS (details of this approach and evaluation can be found in the SI).

## Outlook

We show that the QCL-based DCS may enable VSS measurements that have not been possible using FTIR spectrometers on a routine basis. One such example is VSS measurements of molecules bound within protein active sites, which provide uniquely different environments relative to glass forming solvents. As a result, the Stark response may change if, for instance, the Stark tuning rate is modulated upon substrate or inhibitor binding. Previously, such experiments have predominantly involved vibrational probes in the IR-transparent window, e.g. CO bound to the heme of myoglobin43 or SCN-labelled ketosteroid isomerase[51]. Vibrations in other spectral regions (vibrational probes exist throughout most of the mid-IR[52]) are potentially more difficult to measure and interpret, due to overlap with the background proteins, but can be viable if the linewidths observed upon binding are sufficiently narrow relative to the background transitions, as the Stark signal will be significantly sharper due to the derivative-like nature of the Stark-response. In addition, sufficient signal is often limited by the concentrations achievable for proteins within a frozen glass matrix, which is typically on the order of 1 mM. Therefore, extrapolating from the results of this work, we can see that lowering the solute concentration by 2 orders of magnitude would require an experiment time of 167 h to achieve a Stark spectrum with similar SNR quality as measured during 384 s at 100 mM using FTIR instruments. However, the current DCS setup would be able to reduce this time considerably to below 5 mins (with exact number varying dependent on the chosen SNR evaluation method), enabling such experiments to be routine in the future. Note that these numbers were estimated neglecting any spectral overlap with protein signals in the same spectral region, however, here the increased intensity from a laser source can also offer an advantage.

A second common barrier in VSS is presented by restrictions to duration and amplitude of fields that can be applied to a sample before dielectric breakdown using the established setup with the versatile FTIR. The current implementation using the FTIR is photon-limited due to the globar, and as such the optimal electrode spacing (25 µm) and Ni-coating (4.5 nm) have been chosen to achieve a sufficient SNR for VSS using electric fields of ca. 1 MV/cm on ca. 100 mM sample concentrations. Based on the current experimental constraints with conventional FTIR, we outline how the DCS systems could be used to further optimize the VSS experimental setup. The first advantage of the DCS is the much brighter QCL light source, which could allow for thicker electrode coatings to be used, thereby increasing the robustness and magnitude of electric fields that can be applied (the Stark signal scales with the electric field squared). In addition, a higher brightness allows for adjustments of the optical layout, such as adjusting the light polarization and angle of the incident light with respect to the electrodes (which must be taken into account in determination of the Stark tuning rate) and/or transmitting the light multiple times through the sample (e.g. by replacing one of the two Ni-coated windows with a mirror and extracting light from the window that was used for illumination). Both of these examples could allow for higher signal to be achieved at shorter pathlengths, further enabling optimization of the SNR and electrode spacing to achieve higher applied fields. Finally, the DCS enables the possibility of greatly lowering the required acquisition times to achieve sufficient SNR and reduce the likelihood of dielectric breakdown due to extended application of an electric field to the sample-holder. Taken together, the higher brightness and short acquisition times of the DCS highlight the tremendous promise for increasing the application and feasibility of VSS.

# Conclusions

In this work, we have shown that we can reproduce the Stark spectrum of fluorobenzene using DCS and obtain a Stark tuning rate of $(0.81 \pm 0.09)$ cm$^{-1}$/(MV/cm), which is in agreement with the result from FTIR setup $((0.89 \pm 0.15)$ cm$^{-1}$/(MV/cm)) as well as the reference FTIR value measured in previous work.[45] Using the DCS, we improved the SNR value of the Stark signal almost by an order of magnitude (a factor of 8 using the simplest approach). Furthermore, based on the faster acquisition and brighter light source, we present future applications that may be accessible using DCS. In conclusion, the DCS provides an experimental setup capable of extending the applications of VSS through fast acquisition, short experiment time, and high brightness. Its spectral point spacing of 0.328 cm$^{-1}$ brings additional advantage over conventionally used spectral resolutions.

# Acknowledgements

We would like to thank Chi-Yun Lin for helpful discussions. We thank Prof. Ronald Hanson and Dr. Christopher Strand, as well as Prof. Jonathan Fan and Prof. Steven Boxer of Stanford University for providing instrumentation for experiment execution. We are also grateful for the anonymous Reviewers for very useful inputs.

# Funding

J. K. acknowledges the Deutsche Forschungsgemeinschaft for a Research Fellowship (KO 5464-1/1). S.H.S. and facilities in the Boxer lab are supported by NIH GM118044.

# Conflict of Interest

Authors declare no conflict of interest.

# Supporting Information

1) SNR evaluation

We benchmark the DCS with FTIR, therefore we use SNR as in FTIR using the following formula: $SNR = \frac{Fitted\ Signal_{max} - Fitted\ Signal_{min}}{noise}$, where the numerator corresponds to the signal intensity in the measurement (i.e. the maximum difference signal from the fit described in the manuscript), and $noise$ due to the different spectral characteristics of the two spectrometers, is calculated differently for DCS and FTIR. In DCS, each spectrum consists of a set of points that correspond to the narrow laser lines that are individually detected with their standard deviation. We describe different ways of DCS $noise$ quantification below. For FTIR, $noise$ is equal to the RMSE (root mean squared error), calculated for a linear fit (to zero-line) of the baselined to zero spectral region without spectral features below 1200 cm$^{-1}$. The obtained SNR results are collected in Table 2S. Of note, there are several processes taking place during the experiment, that may influence the spectral response and influence the analysis of the noise – e.g ice formation, nitrogen bubbling, etc. Furthermore, the 2-MeTHF solvent is observed to possess a small Stark contribution in the spectral region analyzed.

Thanks to monitoring of laser line intensity in the DCS system, we can quantify the standard deviation for each spectral point, as presented in the manuscript and visualized in Figure 4 and Figure 7S below. We use the standard deviation values as the base in the following (1.1-1.3) analysis.

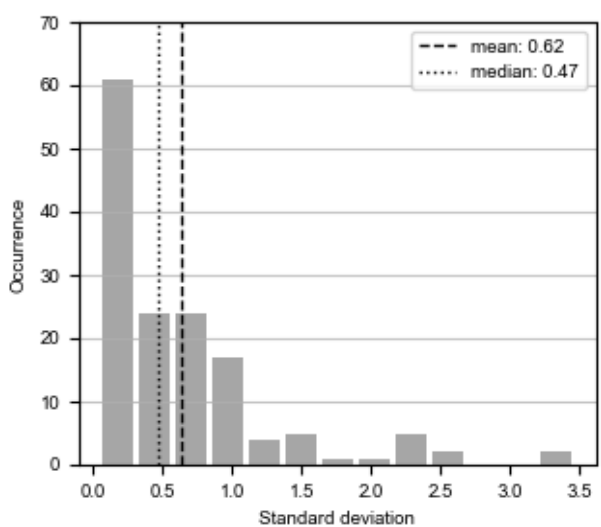

Figure 7S. Histogram of DCS standard deviation values. The analysis is across all spectral elements.

### 1.1. DCS - Mean
The mean value of standard deviation i.e. its arithmetic average, is the simplest way of treating the set, although not the most statistically representative. Nevertheless, this is the value that could be thought of as the most similar to FTIR derived value. The mean of DCS noise is 0.62.

### 1.2. DCS - Median
The median represents how the population is distributed and is more representative of populations that are not normal. The median value for DCS noise is 0.47.

### 1.3. DCS - Mean of (signal over standard deviation)
Another approach can be to calculate SNR at each data point and then take the mean value of it. This will be more affected by very low noise values. A simple arithmetic average of SNR for DCS gives the value of 22.

### 1.4. FTIR
From the fitting of a spectrum to zero (see above) made possible due to using the flat noise characteristics in a spectral region without spectral features, one obtains rmse = 0.32.

### 1.5. Summary
The fitted signal values for DCS and FTIR are 4.53 and 5.04, respectively. For FTIR it results in SNR of ca. 16. FTIR and DCS values for different approaches are collected in Table 2S. To account for the improvement of SNR coming from using the DCS instrument, we are comparing the SNR values in the form of a ratio: $\frac{SNR_{\mathrm{DCS}}}{SNR_{FTIR}}$.

Table 2S. The results of SNR analysis for spectra obtained with DCS and FTIR instruments.

| Approach | $SNR_{DCS}$ | $SNR_{FTIR}$ | $\dfrac{SNR_{DCS}}{SNR_{FTIR}}$ | $\left(\dfrac{SNR_{DCS}}{SNR_{FTIR}}\right)_{corr}$ |
|----------|-------------|--------------|--------------------------------|----------------------------------------------------|
| 1.1 | 7 | 16 | 0.44 | 7.5 |
| 1.2 | 10 | 16 | 0.63 | 10.7 |
| 1.3 | 22 | 16 | 1.4 | 23.6 |

Differences in the values collected in Table 2S highlight the difficulty of assessing SNR for a system that does not have uniformly distributed noise and that has spectral regions with very low noise levels. We decided to use noise level as determined as the median of standard deviation of signal for the DCS system. This value was used to further estimate SNR values for potentially more demanding experiments. The results are presented in Table 3S. From Figure 4 it can be seen that the C-F stretching band is located in the low-noise region of the DCS spectrum. However, it should be noted that this is not the origin of the observed improvement of SNR obtained thanks to DCS system, since the noise was calculated for the whole spectral window covered by DCS. The location of the low noise region only affects the estimated value (the Stark tuning rate) and its uncertainty. As laser development continues, more homogeneous power distributions will be achieved, leading to less wavelength-dependent noise levels throughout the covered spectral range.

Table 3S. Comparison of performances of two setups used in this study. All predictions assume, unless stated otherwise, 100 mM solution of fluorobenzene. SNR were estimated using a square root dependence of the SNR on time and a linear dependence on concentration. The base for SNR estimations are the values from row 1.2 of Table 2S.

| Feature | FTIR | Dual-comb spectrometer |
|---------|------|------------------------|
| SNR per 1 min acquisition time | 6.3 | 62 |
| Acquisition time to obtain SNR = 6.3 | 1 min | 16 ms |
| Acquisition time to get SNR = 6.3 at 1 mM | 167 h | 159 s |
| Noise averaged over the spectrum | ✔ [v] | ✘ |
| Possibility of weighted noise analysis | ✘ | ✔ |

---

[v] only valid when no filters are introduced in the beam path, e.g. to reduce illumination of the detector

2) Representation how the weighted fit was implemented – realized using Python

Requirements: numpy (as np) and scipy.optimize (as optimization).

a. Define the fit formula [separate derivative and pseudo-Voigt formulas not shown]. At first, declare A, mu, sig, and alfa, as the values you found by fitting the absorbance spectrum (a separate step).

```
def fit(x, c1, c2, c3):
    return c1*pVoigt(x, A, mu, sig, alfa)+c2*firstDer(x, A,
mu, sig, alfa)+c3*secondDer(x, A, mu, sig, alfa)
```

b. Fit the data to the sum of derivatives. Inputs: fit – as defined above, WavenumberValues – x axis, AbsorbanceValues – y axis, x0 – initial guess of parameters (here: `x0 = np.array([0, 0, 0])`).

```
params, params_covariance = optimization.curve_fit(fit,
WavenumberValues, AbsorbanceValues, x0, sigma=STD,
absolute_sigma=True)
```
# **sigma=STD** - standard deviation of each laser line intensity; **absolute_sigma=True** refers to the fact that exact values of standard deviation are used, without additional normalization; for FTIR, sigma and absolute_sigma are not specified and used as default. See below.

```
"""sigma : None or M-length sequence or MxM array, optional
    Determines the uncertainty in ydata. If we define
residuals as r = ydata - f(xdata, *popt), then the
interpretation of sigma depends on its number of dimensions:
        A 1-d sigma should contain values of standard
deviations of errors in ydata. In this case, the optimized
function is chisq = sum((r / sigma) ** 2).
        A 2-d sigma should contain the covariance matrix of
errors in ydata. In this case, the optimized function is chisq
= r.T @ inv(sigma) @ r.
        New in version 0.19.
        None (default) is equivalent of 1-d sigma filled with
ones.
    """
```

```
yfitDCS = fit(WavenumberValues, params[0], params[1],
params[2])
```

c. Find the Stark tuning rate and its uncertainty

Calculate the Stark tuning rate:

```
DmuDCS  = np.sqrt(10*params[2])
```

Calculate the uncertainty on the Stark tuning rate:

```
paramsErr = np.sqrt(10*np.sqrt(np.diag(params_covariance)))
```

The Stark tuning rate is DmuDCS +/- paramsErr[2]

    d.  For SNR analysis, one needs to obtain signal value from the fit:

```
sigDCS = max(yfitDCS)-min(yfitDCS)
```
 # signal DCS – from the fit not to rely on noisy data points

   3)  The full FTIR spectrum

The full recorded FTIR spectrum is presented in Figure S.

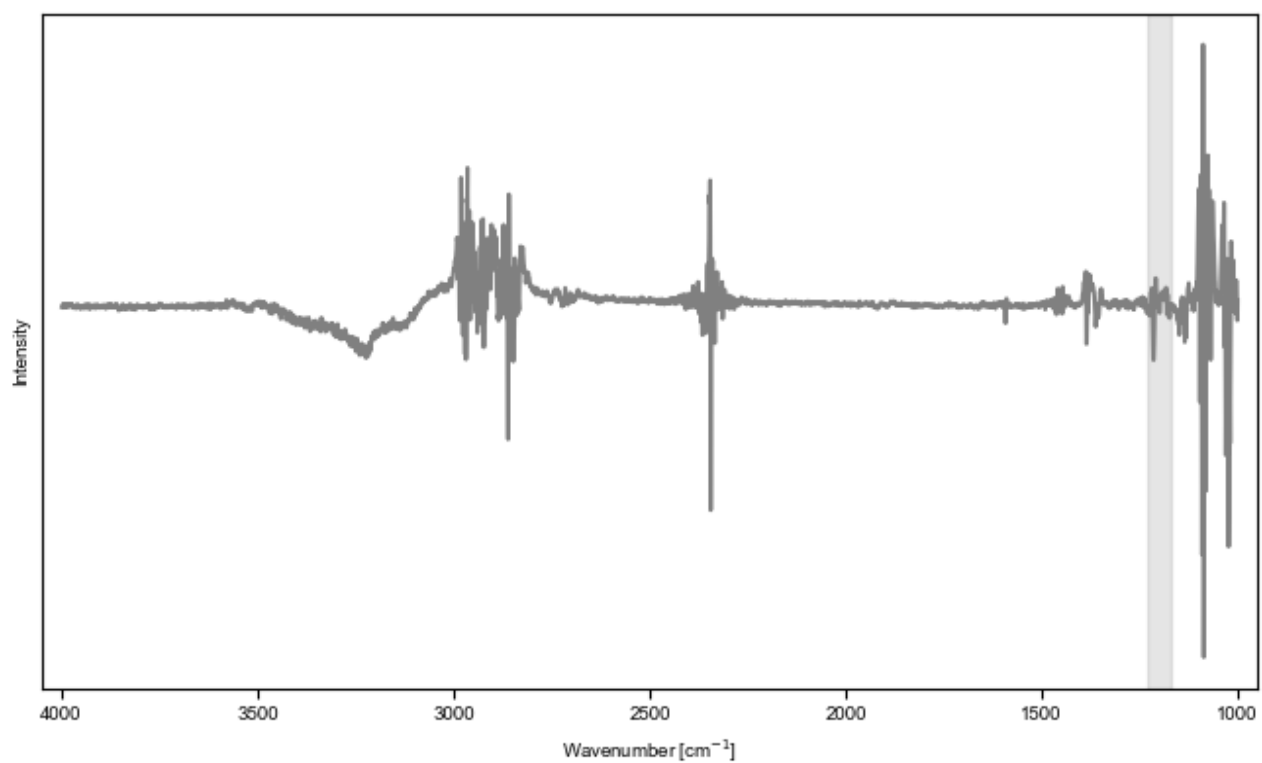

Figure 8S. Full FTIR spectrum. Note features corresponding to e.g. ice formation. Vertical bar corresponds to the window of analysis for the Stark tuning rate determination.